\documentclass[a4paper]{article}
\pdfoutput=1

\usepackage[english]{babel}
\usepackage[utf8x]{inputenc}
\usepackage[T1]{fontenc}
\usepackage{cleveref}
\usepackage{authblk}

\usepackage[a4paper,top=3cm,bottom=2cm,left=3cm,right=3cm,marginparwidth=1.75cm]{geometry}

\usepackage{amsmath}
\usepackage{graphicx}
\usepackage[colorinlistoftodos]{todonotes}
\usepackage[colorlinks=true, allcolors=blue]{hyperref}

\title{Haematopietic stem cells -- \\ entropic landscapes of differentiation}
\author[1,2]{K. Wiesner}
\author[2,3]{J. Teles}
\author[2]{M. Hartnor}
\author[2]{C. Peterson}
\affil[1]{School of Mathematics, University of Bristol, BS8 1TW Bristol, U.K.}
\affil[2]{Computational Biology and Biological Physics, Department of Astronomy and Theoretical Physics, Lund University, 223 62 Lund, Sweden}
\affil[3]{Sainsbury Laboratory, University of Cambridge, CB2 1LR Cambridge, UK}
\date{}                                           

\begin{document}
\maketitle
\begin{abstract}

The metaphor of a potential epigenetic differentiation landscape 
broadly suggests that during differentiation a stem cell follows the steepest descending gradient toward a stable equilibrium state which represents the final cell type. It has been conjectured that there is an analogy to the concept of entropy in statistical mechanics. In this context, in the undifferentiated state the entropy would be large since fewer constraints exist on the gene expression programs of the cell. As differentiation progresses, gene expression programs become more and more constrained and thus the entropy would be expected to decrease. Such an entropy decrease would, in analogy to statistical mechanics, require some form of free energy to decrease accordingly. In order to assess these predictions, we compute the Shannon entropy for time-resolved single-cell gene expression data in two different experimental setups of haematopoietic differentiation. We find that the behaviour of this entropy measure is in contrast to these predictions. In particular, we find that the Shannon entropy is not a decreasing function of developmental pseudo-time but instead it increases toward the point of commitment before decreasing again. This behaviour is consistent with an increase in gene expression disorder observed in populations sampled at the point of commitment. Single cells in these populations exhibit different combinations of regulator activity that suggest the presence of multiple configurations of a potential differentiation network as a result of multiple entry points into the committed state. 
\end{abstract}

\section{Introduction}

The programs governing the function and fate of cells are to a large extent driven by the coordinated activity of transcription factors forming complex and dynamic gene regulatory networks. The activities of transcription factors and other genes involved in cell fate decisions can be measured by a number of different gene expression quantification experiments. Until recently, and due to technical limitations, for a given cell type such experiments had to be done on an ensemble of many cells and, hence, gene expression quantifications represented the average over a given population. This averaging effect hampered the analysis of finer regulatory mechanisms at the single cell level, the fundamental unit for any fate decision process. More recently, a number of novel technologies have facilitated gene expression measurements for individual cells, thereby opening up the possibility of quantifying heterogeneity among cells of a given population and between related populations (for a review see e.g. \cite{Stubbington_2017}. Such heterogeneity could originate from extrinsic factors but also from the intrinsic noise generated by having few copies of molecules involved in transcription and translation. Whether intrinsic noise is simply a result of the stochastic nature of any cellular process or actually plays a mechanistic role in cellular decision-making processes during differentiation is currently the object of intense study.

Entropy in statistical mechanics is a measure of disorder in the macrostate of a system. The more different microstates are visited the higher the entropy. Mathematically, the statistical mechanical entropy  is equivalent to the information-theoretic Shannon entropy, where the latter measures the amount of randomness in a probability distribution \cite{cover2012elements}. Hence, the Shannon entropy of a probability distribution over gene expression levels in a cell population measures the amount of randomness or heterogeneity in its gene expression patterns. Therefore, estimating the Shannon entropy of a cell population might yield insights into the role of gene expression noise, which would be of particular interest in a context of state transitions such as cellular differentiation.

With the upsurge of studies of stem cell commitment processes during the last decade the subject of heterogeneity is of particular interest. Since stem cells and progenitors host the genetic program potential for all mature cell types they can give rise to, one would naively expect them to be strongly disordered in terms of gene expression patterns as compared to the mature cells they originate. Expressing order or disorder as lack thereof by means of entropy could be a way forward in monitoring commitment of stem cells, and differentiation towards mature cells.

We have therefore explored such scenarios of stem cell commitment and differentiation for two haematopoietic differentiation systems: (i) The first system \cite{Guo_2013} consists of long term haematopoietic stem cells (LTHSC) which differentiate into multipotent progenitor (MPP) before bifurcating into common myeloid progenitors (CMP) or common lymphoid progenitors (CLP), as is illustrated in Fig.~\ref{fig.orkin}A. In this first system, we are interested in quantifying the entropy while the system moves from less differentiated to more differentiated compartments and, in particular, in assessing how the entropy behaves before and after the first major branching point. (ii) The second system is an example of a more fine-grained resolution of haematopoietic differentiation. We use gene expression data immediately before and after an erythroid commitment decision \cite{Pina_2012} in the factor-dependent multipotent haematopoietic cell line EML. As in the first system, we are interested in assessing how entropy values change from a less to a more constrained differentiation state, across the time point where an irreversible decision has been made.

\section{Single Cell Gene Expression Data}

For this study, we considered two sets of previously published single-cell quantitative RT-qPCR data that included candidate genes known to be involved at different stages of haematopoietic differentiation. 
From Guo et al. \cite{Guo_2013} we analysed the data from 179 regulators that included lineage-specific transcription factors, epigenetic modifiers and cell-cycle regulators. The expression of these genes was quantified in a total of 191 cells from different stem and progenitor cell populations: long-term haematopoietic stem cells (LT-HSC), multipotent progenitors (MPPs), common lymphoid progenitors (CLP), common myeloid progenitors (CMP), granulocyte-monocyte progenitors (GMP) and megakaryocyte-erythroid progenitors (MEP). For each gene, expression is defined as Log2 expression above the system background $Ct$ of $28$ (i.e. $28$ minus the measured raw $Ct$). $Ct$ values higher than $28$ were transformed to $28$ and defined as being $0$ (no measurable gene expression). For more experimental details on population sorting, PCR protocol and gene filtering/normalization we refer to the original paper \cite{Guo_2013}.
From Pina et al. \cite{Pina_2012}, we analysed single cell gene expression data from different subpopulations of the multipotent haematopoietic cell line EML. More specifically, we focused on RT-qPCR data for 17 genes measured in 319 self-renewing (SR), 109 erythroid-committed (CP) and 83 erythroid-differentiated (Ediff) cells. Through clustering and multivariate methods, the CP population was further subdivided into two compartments, CP1 and CP2, as described in Teles et al. \cite{Teles_2013}. CP1 and CP2 have been inferred to be early and late committed cells, respectively, given the similarity of their gene expression profiles to the SR (in the case of CP1) or Ediff (in the case of CP2) populations. For all genes, expression was originally defined as $\Delta Ct$ for each gene to the reference gene (Atp5a1) and linearly transformed to $\ln(2^{30}-\Delta Ct)$, where $30$ is the experimental detection limit. For more information on culture conditions, cell sorting and gene filtering/normalization we refer to  \cite{Pina_2012}.

\section{Entropy Estimation}

The standard Shannon entropy is a function of a discrete probability distribution while gene expression in general is measured on a continuous scale. Hence, the data need to be discretised for the entropy to be measured. The alternative is to estimate the generalised Shannon entropy for continuous distributions (see, for example \cite{cover2012elements}). However, both definition and estimation of continuous Shannon entropy are afflicted with problems, such as requiring large data and potentially returning negative values. We therefor do not consider the continuous Shannon entropy any further here, but we will offer some insights into its use in the context of gene expression data in a forthcoming publication.

In discretising continuous gene expression data into bins, the decision of how many bins to use is a difficult one when there is no obvious and biologically justified separation between expression levels. Hence, in this study only two obviously separate levels are distinguished between: zero expression level and greater-than-zero expression level. From this, the binary Shannon entropy (Eq.~\ref{eq.entropy}) is estimated. 
The Shannon entropy of a binary probability distribution $P$ over two events (representing the two bins), each  
 with probability $p_0$ and $p_1$, respectively, is defined as:
\begin{align}
\label{eq.entropy}
H(P) :=  - p_0 \log_2(p_0) - p_{1} \log_2(p_1)~,
\end{align}
where $0 \log 0 := 0$. The Shannon entropy is symmetric in the probabilities of the two events, it is zero whenever either $p_0 = 0$ or $p_1 = 0$, and it is maximal when $p_0 = p_1 = 1/2$, in which case $H(P) = 1$. 

The entropies of the gene expression data in this study were estimated using the non-parametric James-Stein-type shrinkage estimator, developed by Hausser and Strimmer \cite{hausser_entropy_2009}. The estimator, together with other entropy estimators, is implemented in the R package `entropy' \cite{R-entropy-package}, which was used here. The James-Stein-type shrinkage estimator is shown to be highly efficient statistically as well as computationally \cite{hausser_entropy_2009}.
Other estimators, such as the maximum-likelihood estimator are known to be very sensitive to even moderately sparse data, which we confirmed in simulations of synthesised data sets (not included here). 

Entropy is not the only measure of randomness or variation of a random variable. An obvious one to compare it to is the variance. In the case of a  binary random variable, there is a straight forward mathematical relation between the variance and the entropy. Using the same notation  as in Eq.~\ref{eq.entropy}, the variance of a binary random variable is given by:

\begin{align}
\text{ Var}(P) = p_1(1-p_1)~.
\end{align}

The variance and the entropy of a binary probability distribution both peak at $p_0 = p_1 = 1/2$ and are equal to zero for $p_0 = 0$ or $p_0 = 1$. Thus, the variance computed for the same data set will show the same qualitative behaviour as the entropy. We computed the sample variance for both gene expression data sets (not included here) and found this mathematical prediction  confirmed. 

The true strength of the Shannon entropy over other statistical measures of randomness is that it can be generalised both to a set of $n$ correlated random variables and that it is an entry point to a whole set of information theoretic tools which quantify randomness of and correlations between any number of variables. Less relevant here but still worth noting is that the Shannon entropy is applicable to data which are non-numeric, such as DNA sequences, molecular configurations, or written text. Furthermore, as mentioned in the beginning, the Shannon entropy is proportional to the statistical mechanical Gibbs entropy (although the debate on the interpretation of this mathematical fact is still ongoing \cite{ladyman2008use}).  Hence,  the Shannon entropy  can be used directly in discussions of a potential epigenetic differentiation landscape 
 imposing  statistical mechanical constraints on genetic development through the laws of thermodynamics. 

 \subsection{Standard error of entropy estimates}
 
To obtain the standard error (the root mean squared error) of the entropy estimates, the non-parametric jackknife method was used \cite{efron_nonparametric_1981}.
 There are many comprehensive expositions of this method,  e.g. \cite{efron1983leisurely, efron_bootstrap_1986}. We briefly summarise it here: for a set of $n$ samples of a random variable (r.v.), an estimator $\hat{\theta}$ of the r.v. (such as the mean, the variance, or the entropy) is computed $n$ times, each time with one of the data points being removed. Call this estimate $\hat{\theta}_{(i)}$, where the $i^{\text th}$ data point was removed. 
 Efron showed  \cite{efron_nonparametric_1981}  that the standard error of the estimate is given by 
 \begin{align}
 \sigma_J = \sqrt{\frac{n-1}{n} \sum_{i=1}^n (\hat{\theta}_{(i)} - \hat{\theta}_{(\cdot)} )^2}~,
 \end{align}
 where $\hat{\theta}_{(\cdot)}$ is the average of the estimates:
 \begin{align}
 \theta_{(\cdot)} = \sum_{i=1}^n \frac{\hat{\theta}_{(i)}}{n}~.
 \end{align}
 

\section{Results}

\subsection{Long term haematopoietic stem cell differentiation }

\begin{figure}[t!]
\begin{center}
\includegraphics[width=.65\textwidth]{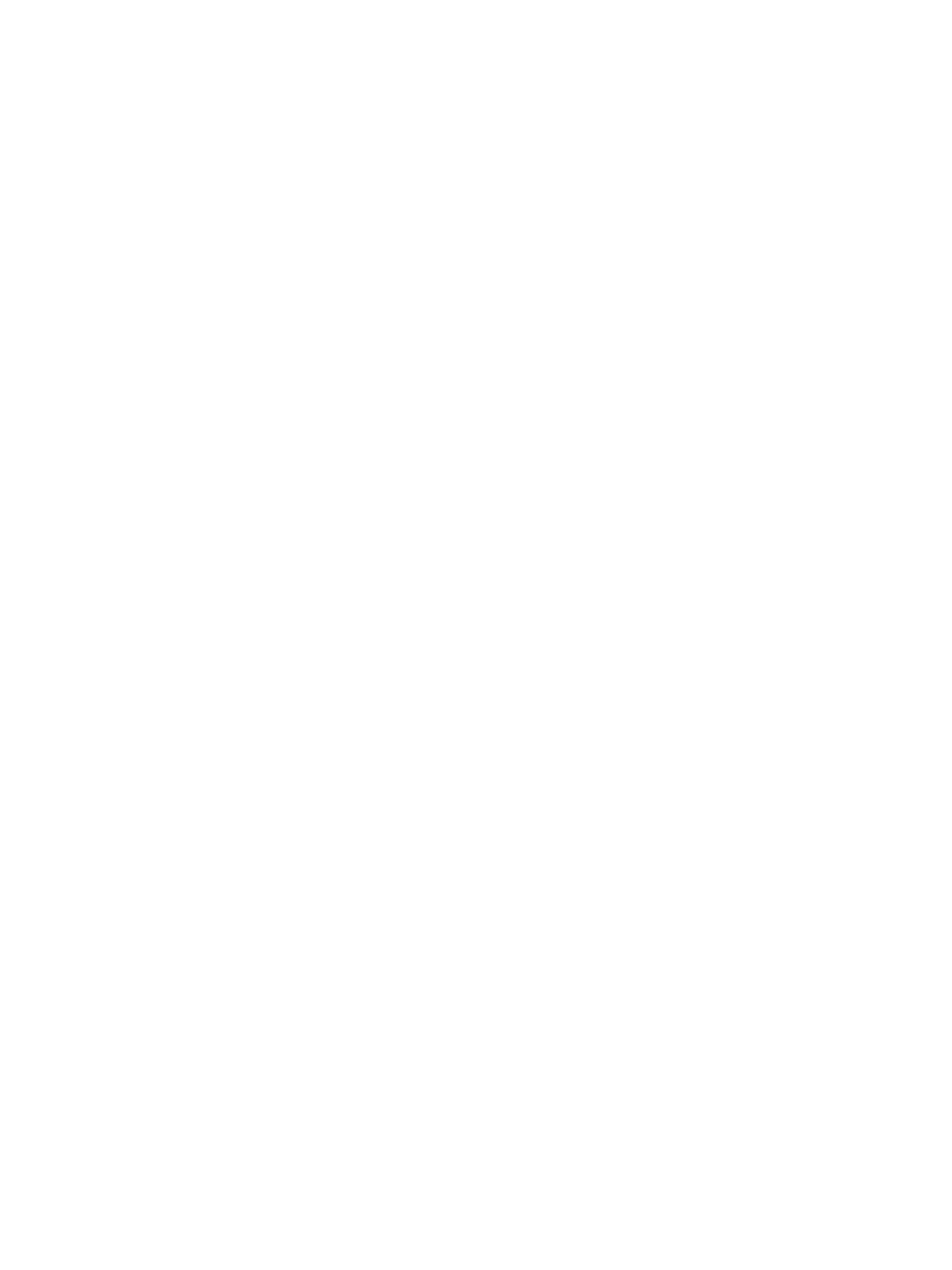}
\caption{\label{fig.orkin} Binary Shannon entropies during haematopoietic differentiation. (A)  Depiction of haematopoietic stem cell differentiation tree. For each of the cellular populations we used single cell gene expression for a number of relevant genes as quantified in \cite{Guo_2013}. 
LTHSC - Long term haematopoietic stem cells; MPP - multipotent progenitors; 
CMP - Common myeloid progenitors; CLP - Common lymphoid progenitors; 
GMP - Granulocyte-monocyte progenitors; MEP - megakaryocyte-erythroid progenitors.
(B) Binary Shannon entropy estimates based on single cell expressions of all genes for each population in (A), with standard error obtained with the jackknife method (see text for details). A significant increase in entropy can be observed immediately after the first branching point, between MPP and CLP/CMP.}
\end{center}
\end{figure}

We estimated the binary Shannon entropy for all cell populations as defined by surface markers of the haematopoietic differentiation tree (Fig.~\ref{fig.orkin}A) described in \cite{Guo_2013} from which the gene expression data is also taken. The results are shown in Fig.~\ref{fig.orkin}

Contrary to what has been conjectured and to what could intuitively a priori be expected, entropy was not found to be a continuously decreasing function along the differentiation pathway (Fig.~\ref{fig.orkin}A). Instead, we observed that entropy slightly decreases from LTHSC to MPP and shows a significant increase between MPP and both CLP and CMP, before decreasing again sharply between CMP and both GMP and MEP. 

\subsection{EML cell line erythroid commitment}

\begin{figure}[t]
\begin{center}
\includegraphics[width=.9\textwidth]{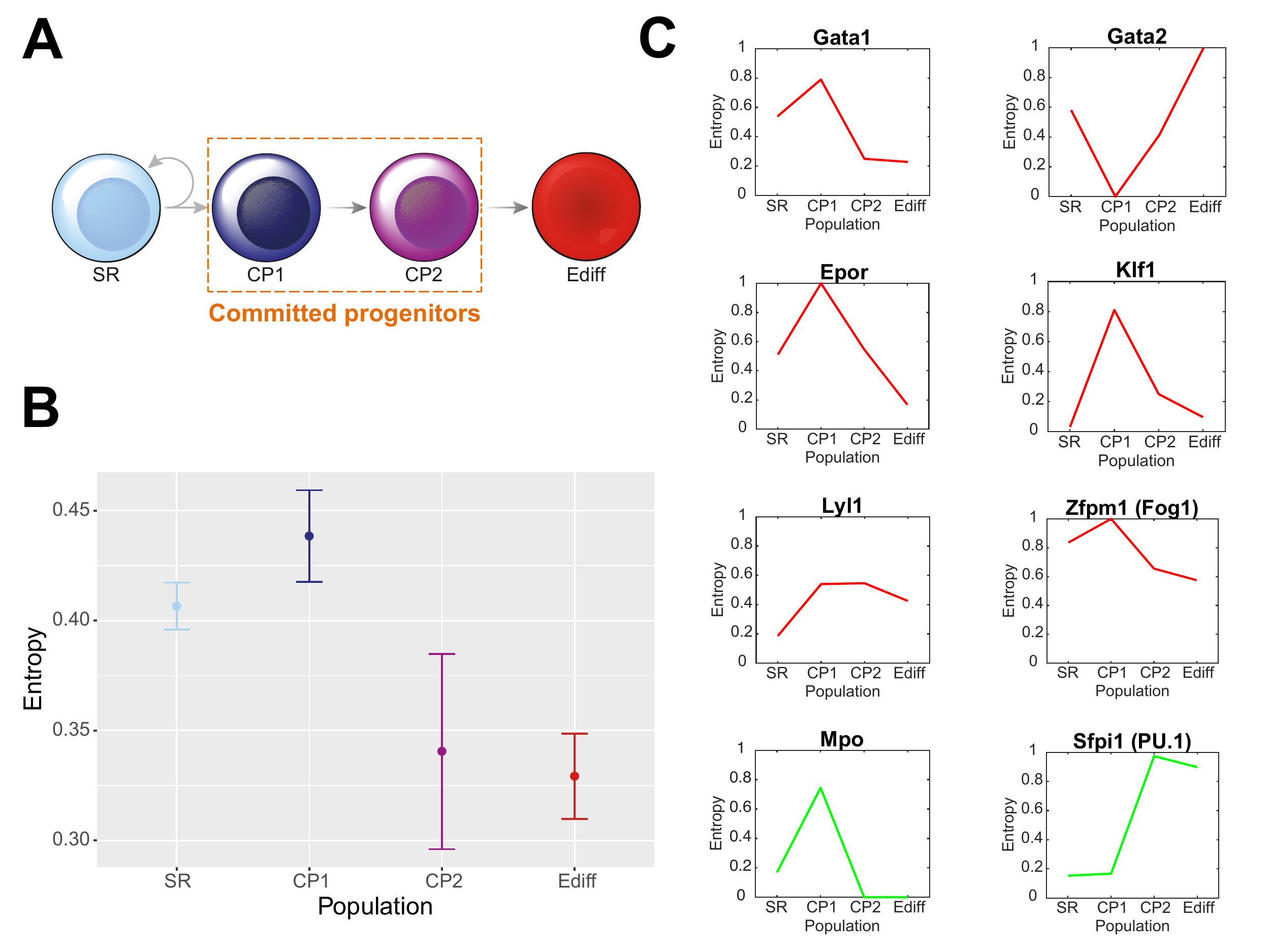}
\caption{\label{fig.eml} Binary Shannon entropies of the EML cell line. (A)  Depiction of subpopulations of the EML cell line allowing the capture of states immediately before (self renewing cells - SR) and after (committed progenitors - CP) commitment. For each population, we used single cell gene expression quantification for a number of candidate genes as measured in \cite{Pina_2012}. CP1 and CP2 are, respectively, early and late committed progenitors; Ediff - Erythroid-differentiated cells. (B) Binary Shannon entropy estimates for all genes in each population in (A), with standard error obtained with jackknife method (see text for details). Entropy values increase immediately after the commitment boundary, in the transition between SR and CP1, decreasing again from CP1 to CP2 and Ediff. (C) Binary Shannon entropy estimates for known genes of interest in erythroid (red) and myeloid (green) differentiation (error bars omitted for simplicity). For the remaining genes in the dataset, please see Supplementary Figure~\ref{fig.supp}.}
\end{center}
\end{figure}

To further investigate entropy dynamics during differentiation, we estimated binary entropies for subpopulations of the EML cell line immediately before and after erythroid commitment, from self-renewing (SR) to committed progenitor (CP) populations (\cite{Pina_2012,Teles_2013}) (Figure~\ref{fig.eml}A). As can be seen in Figure~\ref{fig.eml}B, the entropy values are highest immediately after the decision point, similar to what we observed for the MPP to CMP/CLP transition. Entropy increases from SR to CP1 and decreases again from CP1 to CP2 and from CP2 to Ediff, the late commitment and terminally differentiated populations, respectively.

As previously described by the authors of \cite{Pina_2012,Teles_2013}, CP1 cells show heterogeneity in the expression of known regulators of the erythroid lineage such as Gata1 and Klf1. This observation is consistent with the notion that commitment can be effected even in the absence of the so-called master regulators, and that multiple network configurations can coexist immediately after commitment, subsequently consolidating and becoming more homogeneous in the population as cells differentiate. We tried to further explore this scenario by analyzing the single-gene entropy behaviors for genes involved in erythroid differentiation before and after commitment (i.e. in SR versus CP1 populations). As can be seen  in Figure~\ref{fig.eml}C, Gata1, Zfpm1, Klf1, Epor and Lyl1 all show an increase in entropy from SR to CP1, subsequently decreasing through CP2 and Ediff. Interestingly, myeloid-affiliated genes such as Mpo also show this pattern (PU.1 seems to increase in entropy only in the late commitment CP1 population). Also of note is the fact that Gata2 displays the opposite behavior as the other referred erythroid genes, decreasing in entropy in CP1 to then increase again in CP2 and Ediff.

 \section{Discussion}
 
 \begin{figure}[h]
 \begin{center}
 \includegraphics[width=.8\textwidth]{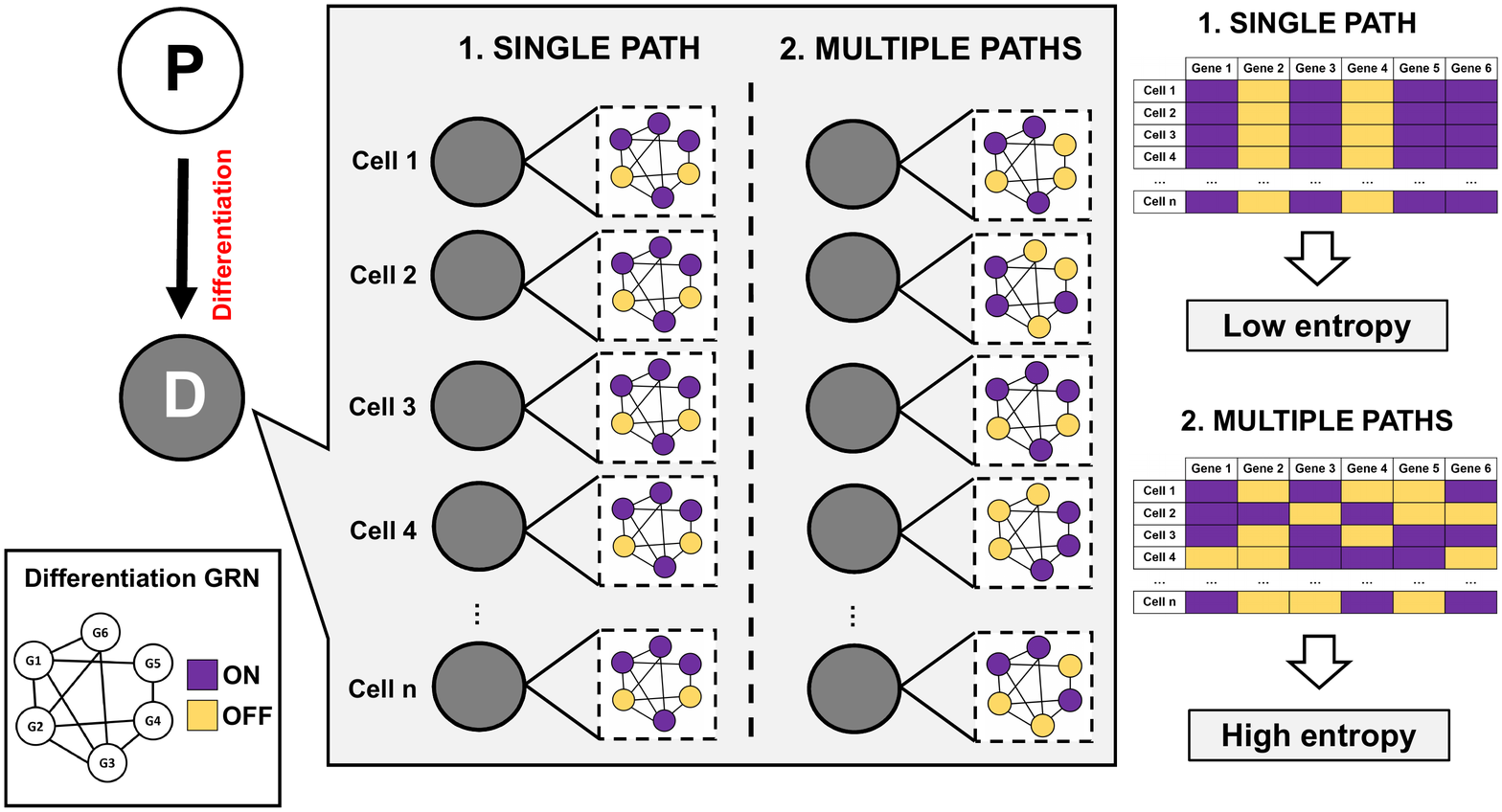}
 \caption{\label{fig.conceptual}Increased binary Shannon entropy in post-commitment cell populations is consistent with multiple paths into the committed state and the co-existence of different states of a differentiation gene regulatory network (GRN). P - progenitor cells; D - differentiated cells. G1 to G6 - gene 1 to gene 6 of a hypothetic differentiation GRN. Purple - Gene is active (ON); Orange - Gene is inactive (OFF).}
 \end{center}
 \end{figure}
 
The interpretation of these results calls for a more careful interpretation of the entropy values and what they may signify in terms of the underlying biology of differentiation (Figure~\ref{fig.conceptual}). Mathematically, a gene has maximum entropy for a given population when half the cells of that population express the gene and the other half does not. High entropy just after a decision point, however, would be, naively, contrary to a more deterministic picture where in order for a cell to progress to a more differentiated state, a set of key regulators would be required to be active and, likewise, key regulators of other lineages that could act as antagonists would need to be repressed. If this assumption was correct, we would expect the entropies of those key regulator genes to be low after a branching point such as the MPP to CMP/CLP transition,  since they would be expected to be either always present or always absent in all post-commitment cells. Since cells can display a high level of heterogeneity in expression of key regulators even after commitment has occurred, this deterministic view is most likely not entirely accurate. These observations suggest that commitment into a more differentiated compartment could thus occur through multiple pathways, each representative of a different substate of the differentiation gene regulatory network (GRN). Higher values of entropy would then be caused by the different expression profiles of these GRN substates when more than one substate is present in the population. 

Our results are consistent with the notion that entropy, as a measure of gene expression disorder, highlights the heterogeneous nature of cell fate decisions through multiple pathways defined by different GRN configurations. In the first analysed dataset, we observed that entropy increases after the MPP branching point, with both CMP and CLP populations showing significantly higher entropy values, as compared to MPP. We further expanded on this observation by analysing a second dataset which sampled populations of the EML cell line, allowing the capture of cellular states immediately before and after the erythroid commitment boundary. As before, we observed an increase in entropy immediately after commitment, from the SR to the CP1 population, consistent with our previous results. Furthermore, we explored the entropy values for single genes and  observed this SR-to-CP1 increase for known erythroid regulators (e.g. Gata1, Klf1 and Fog1) as well as some myeloid regulators (e.g Mpo) (Figure~\ref{fig.eml}C and Supplementary Figure~\ref{fig.supp}). Interestingly, Gata2 shows the opposite trend, with entropy decreasing to zero in the CP1 population, suggesting that for some regulators there is more stringent regulation leading to all cells of the committed population showing the same expression profile (in this case, all cells express Gata2). This result is consistent with previous predictions that Gata2 sets two regulatory modes in SR cells (\cite{Teles_2013}): a restrictive mode when not expressed, effectively blocking commitment, and a permissive mode when expressed, allowing commitment to occur through different combinations of other regulators in the network.  

There are still a number of potential caveats and unresolved questions that require further discussion. An important point is that in both datasets the gene set was chosen a priori and thus results are, by definition, biased. In other words, the entropy behaviour we observe is dependent on the set of genes under analysis. In both systems analysed here, gene selection was informed by potential relevance for the differentiation process, which in principle allows the entropy values to be informative in that context. 

Another question regards the biological interpretation of high entropy values. In the context of the data, and in light of the work of others, we assume the existence of multiple entries into a committed or more differentiated state in which case the interpretation of high entropy is the presence of disorder in the differentiation network, as given by that snapshot of the population (Figure~\ref{fig.conceptual}). An alternative explanation, however, could be that high entropy comes from a gene that is not actively regulated, for instance because it is not important for that population, in which case we would expect a 50/50 presence at any given moment for that population. This is very unlikely if we assume that in order to save energy resources, a cell will most likely not express a gene until it has to do so \cite{wagner2005energy}. In principle, high entropy genes could also be those with cyclic behaviour, e.g. a cell cycle gene. However, such genes are not included in our analysis. 
As a general remark, calculating joint entropies for more than one gene or mutual information values for small sets of genes could allow us to distinguish potentially spurious high entropy values from cases where high entropies are the result of some degree of coordination between genes.

In the first part of our results, we followed the more classical description of the haematopoietic branching tree (Figure~\ref{fig.orkin}A). It should be noted however that this is not a consensual description and multiple versions have been put forward based on different types of data \cite{Ceredig_2009}. Guo et al. suggest that their results support an alternative architecture where lymphomyeloid lineage commitment may happen upstream of the CLP/CMP separation \cite{Adolfsson_2005, Arinobu_2007, Pronk_2007}. In particular, through network inference methods and further validation experiments, they detected signs of coordinated MegE transcriptional priming in HSCs. Using the same set of 179 regulators, our entropy estimates still suggest increased activity at the CLP/CMP bifurcation. 

From the point of view of the data itself, we deliberately use only the binary information of whether gene activity is present or absent. A second aspect of the data is the continuous distribution of values when the gene is active, for which we are currently developing analysis protocols
. From the biological point of view, we can say that in this paper we assume a ``digital'' approach to gene expression where we consider all or nothing effects (the gene is either on or off). This may be a more adequate approximation to some genes than others, where ``analogue'' regulation by fine tuning expression levels may be more relevant. The digital and analogue views are also not mutually exclusive and a more careful exploration of the mechanistic basis and biological function of these two modes would greatly benefit the community \cite{Munsky_2015, Lorberbaum_2013}.

Related work include \cite{MacArthur_2013} where it is argued in general terms that cell population entropy is positively related to developmental potency. In \cite{ridden2015entropy} one also investigates the hypothesis that entropy is monotonically decreasing during differentiation. They develop a Fokker-Planck type model for the expression of a single gene, Sca1, from which they predict a probability density. They compute a differentiation potential which they find to continuously decrease and conclude that the initial density is close to the maximum entropy distribution. In \cite{Teschendorff_2017} the signalling entropy \cite{Gomez-Gardenes_2008} is computed for single cell expression measurements during stem cell differentiation. The main difference to our analysis is in the computation of the entropy. The signalling entropy is extracted from a known protein-protein network whose edges are weighted by the single-cell expression data. This gives rise to a random walk on the network from which entropies are extracted. In contrast to this, our analysis uses the raw expression data directly to compute the entropy of the expression distribution, without the intermediate step of a network. Their results differ from ours as they exhibit monotonic decrease throughout differentiation. In \cite{Richard_2016} a similar entropy analysis was done using a different single-cell data set. A non-
monotonic decrease towards differentiation was found. However, the entropy estimation method is not taking into account the dependency on the number of bins the data are discretised into, which we found to be significant - hence our choice to distinguish between on and off values only. Also, in \cite{Richard_2016}  no comment is made on the statistical accuracy of estimating $N/2$ probabilities from the measurement of $N$ cells. Given the known unreliability of a mere frequency-based estimation of a probability distribution, we hesitate to make more detailed comparisons to our study.


\subsection*{Concluding remarks}

In this study we have found that the Shannon entropy is not a decreasing function of developmental pseudo-time, as predicted by others in the field, but instead it increases toward the point of differentiation before decreasing again. This behaviour was interpreted as  different combinations of regulator activity  suggesting the presence of multiple configurations of the differentiation network as a result of multiple entry points into the committed state. 

What would be the practical use of entropy measurements along a differentiation trajectory? Assuming the interpretation of increased entropy during differentiation is correct, one could measure the entropy in time series or pseudo time-series \cite{Trapnell2014} from static data to obtain a signal for where crucial changes in the development take place, thereby narrowing in on important transitions independent of surface markers.\\

\subsection*{Acknowledgements}
JT and CP were supported by the Swedish Research Council (Vr 621-2008-3074). KW acknowledges additional support through EPSRC grant EP/E501214/1. JT acknowledges additional support through the University of Cambridge/Wellcome Trust ISSF and the Herchel Smith Foundation.

\bibliographystyle{alpha}
\bibliography{HSC}

\newpage
\setcounter{figure}{0}
\renewcommand{\thefigure}{S\arabic{figure}}
\section*{Supplementary Information S1}
 \begin{figure}[h]
 \includegraphics[width=.9\textwidth]{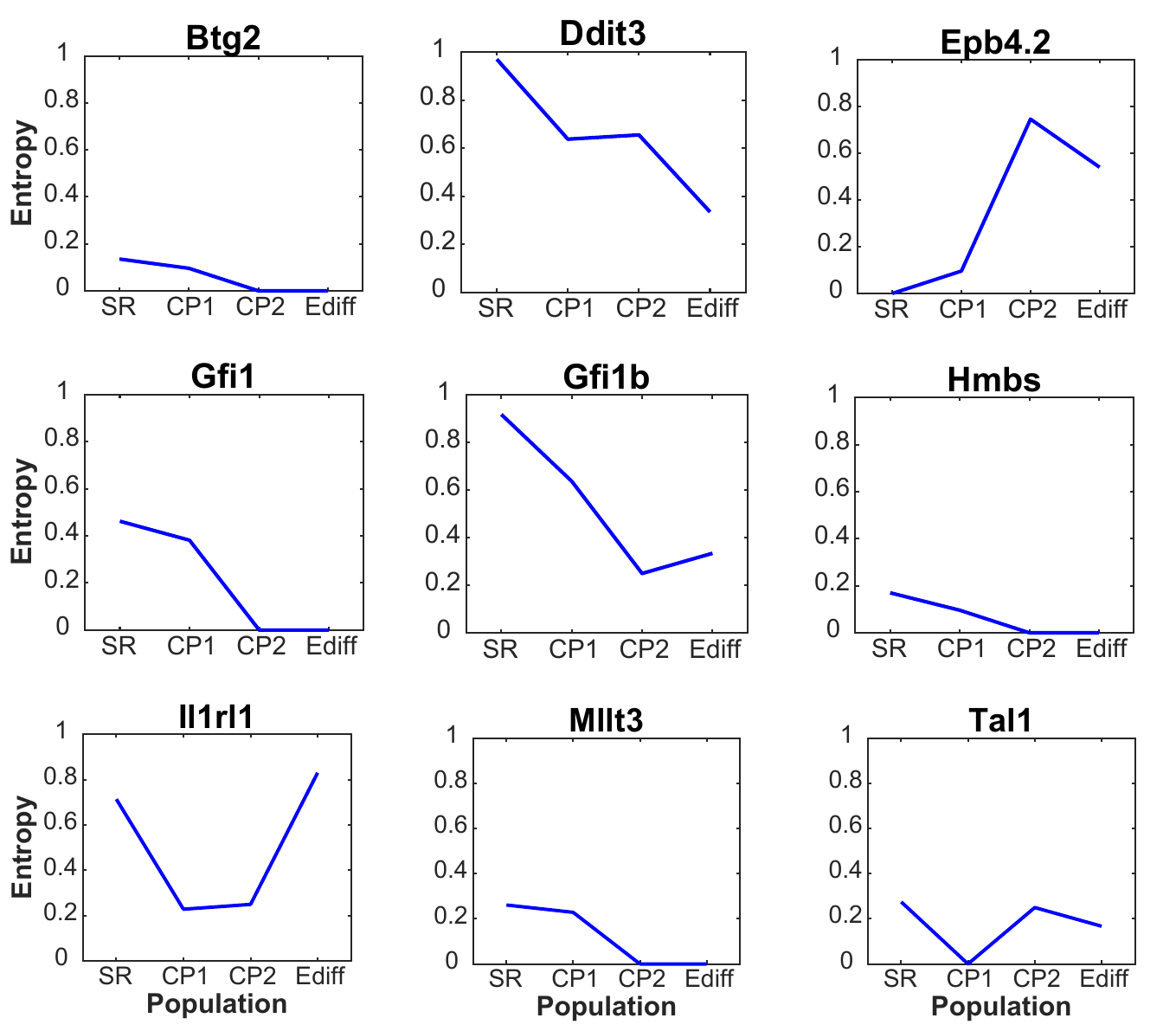}
 \caption{Binary Shannon  entropy for individual genes quantified in different populations of the EML cell line as described in Figure~\ref{fig.eml} in the main text. SR - self-renewing cells; CP1 and CP2 are, respectively, early and late committed progenitors; Ediff - Erythroid-differentiated cells. \label{fig.supp} }
 \end{figure}

\end{document}